# Growth and superconducting properties of Cd-doped La(O,F)BiS$_2$ single crystals

Masanori Nagao[a,c,*,1], Masashi Tanaka[b,c,**,1], Satoshi Watauchi[a], Yoshihiko Takano[c], and Isao Tanaka[a]

[a]*University of Yamanashi, 7-32 Miyamae, Kofu, Yamanashi 400-8511, Japan*

[b]*Graduate School of Engineering, Kyushu Institute of Technology, 1-1 Sensui-cho, Tobata, Kitakyushu, Fukuoka 804-8550, Japan*

[c]*MANA National Institute for Materials Science, 1-2-1 Sengen, Tsukuba, Ibaraki 305-0047, Japan*

Corresponding Authors

*Masanori Nagao for the correspondence of crystal growth and the compositional analysis

Postal address: Center for Crystal Science and Technology, University of Yamanashi

Miyamae 7-32, Kofu, Yamanashi 400-8511, Japan

Telephone number: (+81)55-220-8610

Fax number: (+81)55-254-3035

E-mail address: mnagao@yamanashi.ac.jp

**Masashi Tanaka for the correspondence of crystallographic analysis

Postal address: Graduate School of Engineering, Kyushu Institute of Technology

1-1 Sensui-cho, Tobata, Kitakyushu, Fukuoka 804-8550, Japan

Telephone & Fax number: (+81)93-884-3204

E-mail address: mtanaka@mns.kyutech.ac.jp

Author Contributions

[1] M. Nagao and M. Tanaka contributed equally.

**Abstract**

Cd-doped La(O,F)BiS$_2$ single crystals were grown using a CsCl/KCl flux. The grown crystals have a plate-like shape with ~1-2 mm square size in a well-developed *ab*-plane. The Cd doping in the crystals was successfully characterized by single crystal X-ray diffraction and electron probe microanalysis techniques. The superconductivity of La(O,F)BiS$_2$ was gradually suppressed with Cd doping. The superconducting transition temperature with zero resistivity of La(O$_{0.54}$F$_{0.46}$)(Bi$_{0.92}$Cd$_{0.08}$)S$_2$ was 2.3 K. The Cd doping does not change the superconducting anisotropy so much, albeit the considerable suppression of $T_c$.

**Main text**

**1. Introduction**

BiS$_2$-layered superconductors provide a variety of compounds, such as Bi$_4$O$_4$S$_3$ [1], $R$(O,F)BiS$_2$ ($R$: La, Ce, Pr, Nd, Yb, Bi) [2-8], $A$FBiS$_2$ ($A$: Sr, Eu) [9-11], La$_{1-u}$M$_u$OBiS$_2$ ($M$: Ti, Zr, Hf, Th) [12], and Eu$_3$F$_4$Bi$_2$S$_4$ [13]. The S-site in these compounds can be substituted by Se (BiSe$_2$- and BiSSe-based compounds) [14,15], where these compounds are now called Bi$Ch_2$-based compounds ($Ch$: S, Se). The Bi-site also can be replaced by Sb, as the example of $R$(O,F)SbS$_2$. However, SbS$_2$-based compounds do not show superconductivity [16]. On the other hand, the partial substitution of Ag [17], Cu [18], and Pb [19] for Bi-site all show superconductivity. Especially, Pb substitution enhances the superconducting transition temperature ($T_c$) [19]. These results indicate that substitution of the other elements for Bi-site may enhance superconducting properties. In the view point of the ionic radii, the candidates of substitution species are Hg, Tl, Cd, and so on. It has been reported that it is able to substitute Cd for Bi-sites in the cuprate superconductors, Bi$_2$Sr$_2$CaCu$_2$O$_{8+d}$ [20]. Then we focused on the doping effect of Cd for Bi-site.

In this paper, we synthesized the Cd-doped La(O,F)BiS$_2$ single crystals by the CsCl/KCl flux method [21-23]. The grown single crystals were characterized by single crystal X-ray structural analysis, and investigated its doping effect on the superconducting properties including the anisotropy.

## 2. Experimental

Cd-doped La(O,F)BiS$_2$ single crystals were grown by a high-temperature flux method in an evacuated quartz tube using La$_2$S$_3$, Cd, Bi, S, Bi$_2$S$_3$, Bi$_2$O$_3$, BiF$_3$, CsCl, and KCl as raw materials. The synthesis and characterization procedures are similar to the previous studies in the literature, except for the Cd-doping [22]. The raw materials were weighed with the nominal composition of La(O$_{0.5}$F$_{0.5}$)(Bi$_{1-x}$Cd$_x$)S$_2$ ($x$ = 0-0.20). The mixture of the raw materials (0.8 g) and the CsCl/KCl flux (5.0 g) with a molar ratio of 5:3 was combined using a mortar, and then sealed into a quartz tube under vacuum (~10 Pa). This mixed powder in the quartz tube was heated at 800 °C for 10 h, cooled slowly down to 600 °C at a rate of 1 °C/h, and then furnace-cooled to room temperature. The quartz tube was opened in air, and the flux in the obtained material was dissolved into distilled water. The remained product was then filtered, and washed with distilled water.

Powder X-ray diffraction (XRD) patterns were measured by using Rigaku MultiFlex with CuK$\alpha$ radiation. Single crystal XRD structural analysis was carried out using a Rigaku Mercury CCD diffractometer with graphite monochromated MoK$\alpha$ radiation ($\lambda$ = 0.71072 Å) (Rigaku, XtalLABmini). The crystal structure was solved and refined by using the program SHELXT and SHELXL [24,25], respectively, in the WinGX software package [26]. The compositional ratio of the single crystals was evaluated by energy dispersive spectroscopic (EDS) analysis and electron probe microanalysis (EPMA) associated with the observation of the microstructure by using scanning electron microscopy (SEM). The magnetization–temperature ($M$–$T$) curves under zero-field cooling (ZFC) were measured using a superconducting quantum interface device (SQUID) with an applied magnetic field of 10 Oe parallel to the $c$-axis, except for a sample with $x$ = 0.10 (100 Oe). The

resistivity–temperature ($\rho$–$T$) characteristics were measured by the standard four-probe method with a constant current density mode using physical property measurement system (Quantum Design; PPMS DynaCool). The electrical terminals were fabricated by Ag paste. The angular ($\theta$) dependence of resistivity ($\rho$) in the flux liquid state was measured under various magnetic fields ($H$). The superconducting anisotropy ($\gamma_s$) is estimated using the effective mass model [27].

## 3. Results and discussion

Obtained Cd-doped La(O,F)BiS$_2$ single crystals are plate-like shape with ~1-2 mm in size with a thickness of 10-50 μm. Figure 1 shows XRD patterns of a well-developed plane in a single crystal grown from starting powder with nominal composition of La(O$_{0.50}$F$_{0.50}$)(Bi$_{1-x}$Cd$_x$)S$_2$. The XRD patterns of the grown single crystals with $x$ = 0, 0.05, 0.10 show only 00$l$ diffraction peaks of La(O,F)BiS$_2$ structure, indicating that the $ab$-plane is well-developed. The typical back scattered electron (BSE) image of obtained single crystals is shown in Fig.2 (a). However, when $x$ is increased up to 0.15, the XRD pattern shows multiple phase of La(O,F)BiS$_2$ and Cs$_2$Bi$_2$CdS$_5$ [28]. This is also clearly seen in a BSE image with two kinds of contrasted regions as shown in Fig. 2 (b). The EDS compositional analysis indicates that the blight and dark contrasted regions correspond to Cs$_2$Bi$_2$CdS$_5$ and Cd-doped La(O,F)BiS$_2$, respectively. XRD pattern of the single crystals with $x$ = 0.20 (the top line of Fig.1) was identified to that of Cs$_2$Bi$_2$CdS$_5$ with a well-developed plane of (101). La(O,F)BiS$_2$ have never grown in the nominal composition of $x \geq 0.20$ within the growth condition of this study. These results suggest that the Cs$_2$Bi$_2$CdS$_5$ starts to grow above $x$ = 0.15.

In the obtained single crystals with $x$ = 0-0.10, they were composed of La, O, F, Bi, Cd, and S

elements, whereas the flux components of Cs, K, and Cl elements were not detected by qualitative analysis of EPMA within the detection limit of 0.1 wt%. The estimated Cd-contents ($y$), estimated F-contents ($z$), and the $c$-axis lattice parameter ($c$) in the obtained single crystals are summarized in Table I. The values of $x$ and $y$ were normalized by the total Bi+Cd amount, and the values of $z$ were also normalized by the total O+F amount. The value of $y$ increased up to 0.12 with increasing $x$ to 0.10. The $c$-axis lattice parameters of the obtained single crystals decreased with increase of $y$, but the values of $z$ are comparable to each other. If we assume that the $c$-axis lattice parameter is correlated to the both of the estimated Cd-contents ($y$) and F-contents ($z$), these results suggest that Cd is doped in the obtained single crystals, and the solubility limit of Cd in La(O,F)BiS$_2$ is ca. 0.12.

The single crystal structural analysis was successfully performed using the crystals with $y$ = 0.08. The refinement converged to the $R1$ value of 7.19 % for $I \geq 2\sigma(I)$ for 16 variables including anisotropic parameters. The atomic coordinates and displacement parameters are given in Tables II and III. The compound crystallizes with space group $P4/nmm$ (lattice parameters $a$ = 4.0647(7) Å, $c$ = 13.328(3) Å, and $Z$ = 1). The basic structure of this crystal is isostructural with La(O,F)BiS$_2$, La(O,F)BiSSe and La(O,F)BiSe$_2$ [2,14,15]. All atoms were located at special positions and their anisotropic displacement factors were all positive and within similar ranges, all of which appeared to be physically reasonable. Since the occupancy refinement in O and F under constrain of Occ.(O) + Occ.(F) = 1 did not converge in the refinement, the occupancies were fixed to the EPMA-estimated composition, O:F = 0.54:0.46. Note that the refinement of Cd inclusion was converged only in the case of partial substitution of the Bi site with reasonable occupation value. Taking into account the result of EPMA analysis, Cd atoms are successfully doped to the Bi sites. Figure 3 shows the $M$–$T$ curves for the Cd-doped and non-doped La(O,F)BiS$_2$ single crystals with comparable F-contents. The shielding volume fraction was strongly suppressed by Cd doping, compared to the non Cd-doped La(O,F)BiS$_2$. The superconducting transition was not observed in the single crystals with $y$ = 0.12 down to 1.8 K. Figure 4 shows the $\rho$–$T$ characteristics down to 1.8 K in the single crystals of the Cd-doped and non-doped La(O,F)BiS$_2$ with comparable F-contents ($z$ = 0.46) [22]. The $\rho$–$T$ characteristics of the Cd-doped La(O,F)BiS$_2$ single crystals with $y$ = 0.12 showed semiconducting-like behavior, and it did not show zero resistivity down to 1.8 K. The Cd-doped La(O,F)BiS$_2$ single crystals with $y$ = 0.08 exhibit a resistivity drop at around 4 K, and zero resistivity below 2.3 K. These results indicate that the temperature of zero resistivity was also suppressed by Cd doping, compared to that found in the single crystals of non Cd-doped La(O,F)BiS$_2$ with comparable F-contents. Impurity doping into the superconducting layer usually leads to degradation of superconductivity in the layered superconductor. Cd-doping in this study may also affect to the superconducting properties, then the superconductivity in this compound would lead to suppression. As shown in the inset of figure 4, the resistivity of Cd-doped La(O,F)BiS$_2$ single crystal with $y$ = 0.08 shows two superconducting onsets around 4 K and 3 K. This is probably attributed to the tiny inhomogeneous distribution of Cd elements in the obtained single crystals.

The resistivity ($\rho$) was measured under different magnetic fields ($H$) in the flux liquid state to estimate the superconducting anisotropy ($\gamma_s$) by using the same method with Refs. 29 and 30. The reduced field ($H_{red}$) is calculated using the following equation for an effective mass model:

$$H_{red} = H(\sin^2\theta + \gamma_s^{-2}\cos^2\theta)^{1/2} \qquad (1)$$

where $\theta$ is the angle between the $ab$-plane and the magnetic field [27]. The $\gamma_s$ was estimated from the best scaling of the $\rho$–$H_{red}$ relationship. Figure 5 shows the $\theta$ dependence of $\rho$ for different magnetic fields ($H$ = 0.1-9.0 T) in the flux liquid state for Cd-doped La(O,F)BiS$_2$ single crystals with $y$ = 0.08. The $\rho$–$\theta$ curve exhibited a two-fold symmetry. Figure 6 shows the $\rho$–$H_{red}$ scaling obtained from the $\rho$–$\theta$ curves in figure 5 using Eq. (1). The scaling was performed by taking $\gamma_s$ = 36. The value of $\gamma_s$ in Cd-doped La(O,F)BiS$_2$ single crystals with $y$ = 0.08 was almost the same compared to that of non-doped La(O,F)BiS$_2$ single crystals with comparable F-contents ($z$ = 0.46) ($\gamma_s$ = 23-37) [31]. These results suggest that the Cd doping does not change the anisotropy so much albeit the strong suppression of $T_c$.

## 4. Conclusion

Cd-doped La(O,F)BiS$_2$ single crystals were grown by using CsCl/KCl flux method with ~1-2 mm square size in a well-developed *ab*-plane. The X-ray single crystal structural refinement and EPMA compositional analysis were well revealed the Cd doping to the single crystals. The solubility limit of Cd to Bi-site in La(O,F)BiS$_2$ single crystals was found to be around 12 at%. The temperature of zero resistivity of the obtained single crystals was gradually suppressed with Cd-doping. The superconducting transition temperature with zero resistivity of La(O$_{0.54}$F$_{0.46}$)(Bi$_{0.92}$Cd$_{0.08}$)S$_2$ was 2.3 K, and the superconducting anisotropy was 36. The suppression mechanism of $T_c$ unlike Pb-doped BiS$_2$-based compounds should be investigated in the future, by using the other method, such as scanning tunneling spectroscopy, photoemission spectroscopy and so on.

**Acknowledgments**

The authors would like to thank Dr. A. Miura (Hokkaido University) for useful discussion and critical reading. This work was supported by JSPS KAKENHI (Grant-in-Aid for challenging Exploratory Research) Grant No. 15K14113, and partially supported by JST CREST Grant No. JPMJCR16Q6.

**Supplementary material**

Supporting Information: CIF file for Cd-doped La(O,F)BiS$_2$.

**Figure captions**

Figure 1. XRD pattern of well-developed plane of the single crystals grown from the starting powder with nominal composition of La($O_{0.5}F_{0.5}$)($Bi_{1-x}Cd_x$)$S_2$ ($x$ = 0-0.20).

Figure 2. BSE image of the obtained single crystals with (a) $x$ = 0.05, (b) $x$ = 0.15.

Figure 3. $M$–$T$ curves of the Cd-doped and non-doped La(O,F)BiS$_2$ single crystals with comparable F-contents ($z$ = 0.46).

Figure 4. $\rho$–$T$ characteristics along the *ab*-plane of the Cd-doped and non-doped La(O,F)BiS$_2$ single crystals with comparable F-contents ($z$ = 0.46). The inset is an enlargement of the superconducting transition.

Figure 5. Angular ($\theta$) dependence of resistivity ($\rho$) in flux liquid state at various magnetic fields for the single crystal of Cd-doped La(O,F)BiS$_2$ with $y$ = 0.08.

Figure 6. Scaling of angular ($\theta$) dependence of resistivity ($\rho$) at a reduced magnetic field $H_{red}$ = $H(\sin^2\theta + \gamma_s^{-2}\cos^2\theta)^{1/2}$. $H_{red}$ is calculated from the same data set in figure 5.

Table I. Nominal Cd-content ($x$), the estimated Cd-content ($y$), estimated F-content ($z$) and $c$-axis lattice parameter ($c$) in the obtained single crystals. The Cd-content $y$ and F-content $z$ were normalized by total Bi+Cd and O+F amounts, respectively.

|  | Nominal Cd-content ($x$) in La(O$_{0.5}$F$_{0.5}$)(Bi$_{1-x}$Cd$_x$)S$_2$ | | |
|---|---|---|---|
|  | 0 | 0.05 | 0.10 |
| Estimated Cd-content ($y$) | 0 | 0.08 | 0.12 |
| Estimated F-content ($z$) | 0.43 | 0.46 | 0.42 |
| $c$-axis lattice parameter ($c$) | 13.42 (Å) | 13.35 (Å) | 13.28 (Å) |

Table II. Crystallographic data for the Cd-doped La(O,F)BiS$_2$.

| | |
|---|---|
| Structural formula | La$_2$O$_{1.08}$F$_{0.92}$Bi$_{1.82}$Cd$_{0.18}$S$_4$ |
| Formula weight | 841.40 |
| Crystal dimensions (mm) | 0.12 x 0.05 x 0.02 |
| Crystal shape | Platelet |
| Crystal system | Tetragonal |
| Space group | $P4/nmm$(No. 129) |
| $a$ (Å) | 4.0647(7) |
| $c$ (Å) | 13.328(3) |
| $V$ (Å$^3$) | 220.20(9) |
| $Z$ | 1 |
| $d_{calc}$ (g/cm$^3$) | 6.345 |
| Temperature (K) | 293 |
| $\lambda$ (Å) | 0.71073 (MoK$\alpha$) |
| $\mu$ (mm$^{-1}$) | 17.869 |
| Absorption correction | Empirical |
| $\theta_{max}$ (°) | 27.492 |
| Index ranges | -5<$h$<5, -5<$k$<5, -17<$l$<17 |
| Total reflections | 2063 |
| Unique reflections | 192 |
| Observed [$I \geq 2\sigma(I)$] | 166 |
| $R_{int}$ for all reflections | 0.3936 |
| No. of variables | 16 |
| $R1/wR2$ [$I \geq 2\sigma(I)$] | 0.0719/0.1604 |
| $R1/wR2$ (all data) | 0.0888/0.1648 |
| GOF on $F_o^2$ | 1.027 |
| Max./Min. residual density (e$^-$/Å$^3$) | 3.93 / -6.79 |

Table III. Atomic coordinates, atomic displacement parameters (Å$^2$) for the Cd-doped La(O,F)BiS$_2$.

| Site | Wyck. | S.O.F | x/a | y/b | z/c | $U_{11}$ | $U_{22}$ | $U_{33}$ | $U_{eq}$ |
|---|---|---|---|---|---|---|---|---|---|
| La | 2c | 1 | 3/4 | 3/4 | 0.1010(2) | 0.0074(13) | 0.0074(13) | 0.026(2) | 0.0135(13) |
| Bi/Cd | 2c | 0.91(3)/0.09(3) | 1/4 | 1/4 | 0.37862(14) | 0.0105(10) | 0.0105(10) | 0.0205(14) | 0.0138(9) |
| S1 | 2c | 1 | 1/4 | 1/4 | 0.1891(8) | 0.008(4) | 0.008(4) | 0.014(6) | 0.010(3) |
| S2 | 2c | 1 | 3/4 | 3/4 | 0.3776(11) | 0.023(6) | 0.023(6) | 0.040(9) | 0.029(4) |
| O/F | 2a | 0.54/0.46(Fix) | 1/4 | 3/4 | 0 | 0.002(9) | 0.002(9) | 0.04(2) | 0.016(7) |

Note: $U_{12}$, $U_{13}$, and $U_{23}$ are 0, and $U_{eq}$ is defined as one-third of the trace of the orthogonalized $U$ tensor.

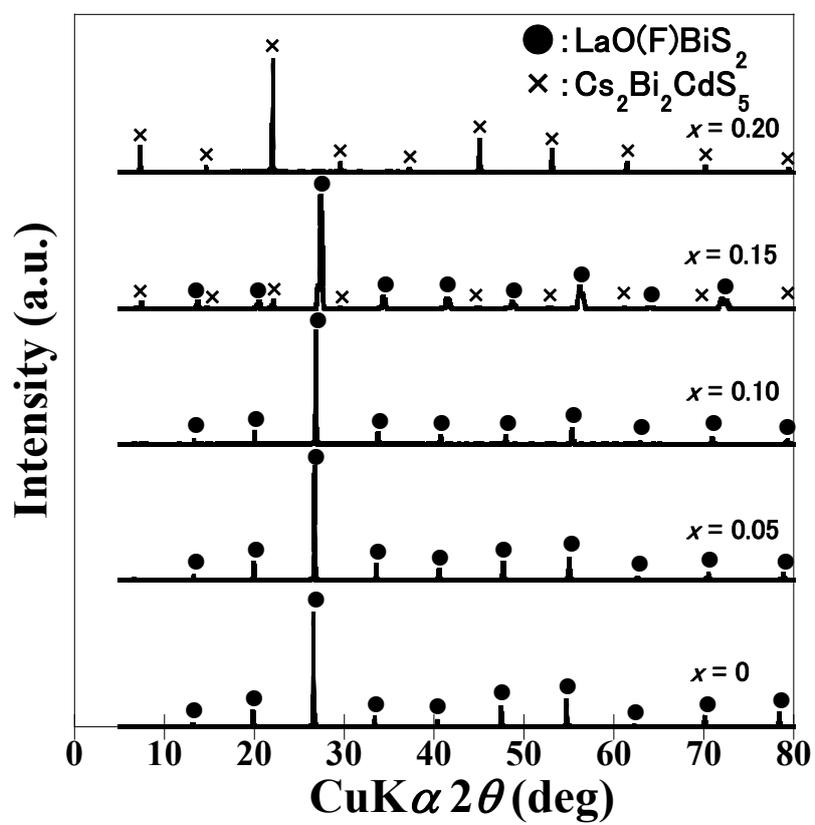

**Figure 1**

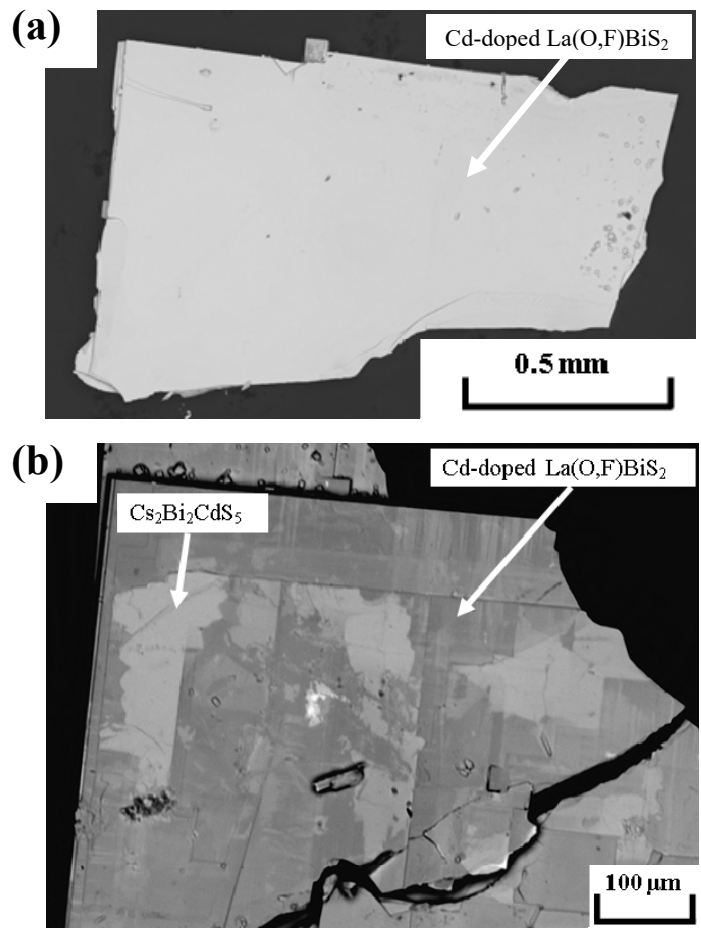

**Figure 2**

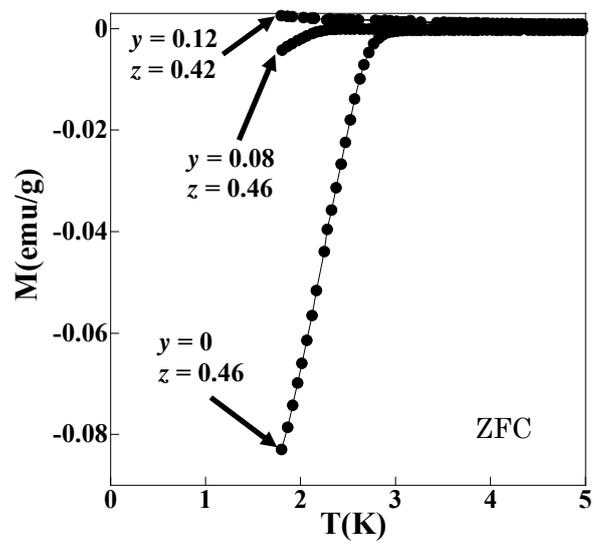

**Figure 3**

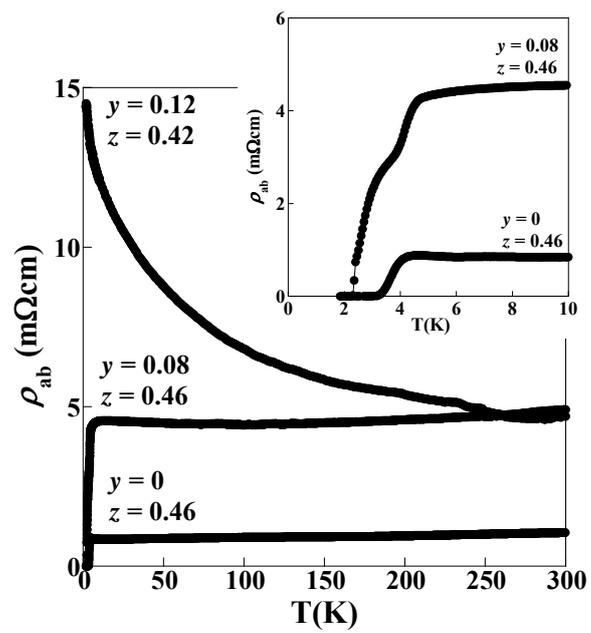

**Figure 4**

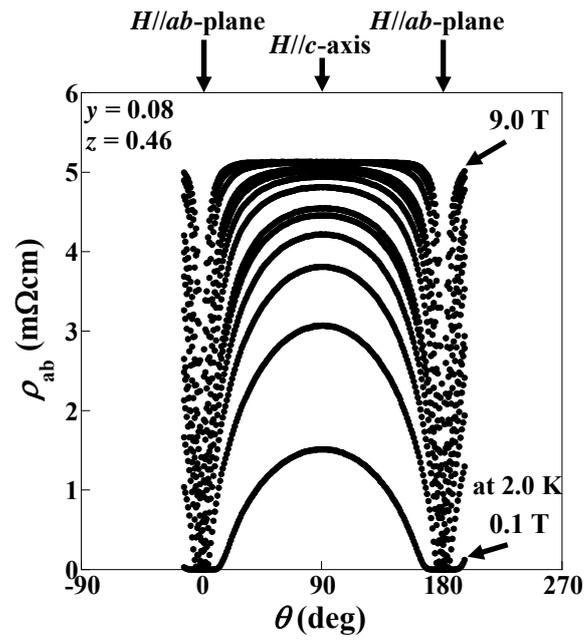

**Figure 5**

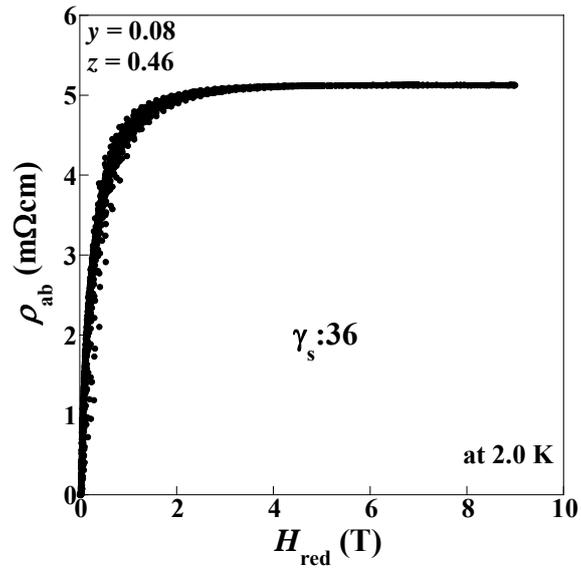

**Figure 6**